\documentclass[epj,twocolumn]{webofc}
\usepackage[varg]{txfonts}   
\woctitle{RPSCINT2013}
\begin{document}
\def\LEO{Li$_6$Eu(BO$_3$)$_3$ }
\def\LMO{Li$_2$MoO$_4$ }

\title{Li-containing scintillating bolometers for low background physics}
%
%

\author{L. Pattavina\inst{1}\fnsep\thanks{\email{luca.pattavina@lngs.infn.it}}
        for the LUCIFER collaboration
}

\institute{INFN - Laboratori Nazionali del Gran Sasso, I-67010 Assergi (AQ) - Italy}

\abstract{%
We present the performances of Li-based compounds used as scintillating bolometer for rare decay studies such as double-beta decay and direct dark matter investigations. The compounds are tested in a dilution refrigerator installed in the underground laboratory of Laboratori Nazionali del Gran Sasso (Italy). Low temperature scintillating properties are investigated by means of different radioactive sources, and the radio-purity level for internal contaminations are estimated for possible employment for next generation experiments.}
\maketitle

\section{Introduction}
\label{intro}

Lithium containing crystal scintillators (such as \LMO, LiF and \LEO) can have important applications in applied physics
(detection of neutron fluxes through $n$ captures by $^6$Li) as well as in fundamental physics in the investigation of neutrinoless double-beta decay (DBD), dark matter (DM) and solar $^7$Li axions.\newline

Neutron induced nuclear recoils can mimic the WIMP signal in DM direct searches, so the evaluation and the minimization of the neutron background is mandatory in order to reach elusive rates in the detectors. The expected fluxes in underground laboratories are very weak ($<$10$^6$ n/cm$^2$/s for neutrons coming form natural radioactivity and $<$10$^9$ n/cm$^2$/s for $\mu$-induced neutrons~\cite{neutrons}) and can be strongly reduced with active and passive shields. Among the various techniques for the estimation of very low neutron backgrounds and their energy distributions the most promising one are based on neutron-induced nuclear reactions: the neutron energy is shared between the reaction products and the estimation of the deposited energy gives directly the measurement of the incident neutron energy. $^6$Li is a favorable isotope for this type of investigation due to its large absorption neutron cross-section in a wide energy range (from few eV up to several MeV).\newline

Double-beta decay is a rare nuclear process where it occurs the simultaneous emission of two electrons from the nucleus $(A,Z)$ $\to$ $(A,Z\pm2)$. Two neutrinos DBD, when two neutrinos are also emitted, is allowed in the Standard Model (SM) of particle physics, measured half-lives range from
$T_{1/2}$ $\simeq$ $10^{18}-10^{24}$ yr \cite{Ver12}. Neutrinoless DBD is forbidden in SM because it violates the lepton number conservation by two units; however, it is predicted by many SM extensions which describes neutrino as Majorana particle ($\nu=\overline{\nu}$) with non-zero mass.
$^{100}$Mo is one of the favorite isotopes in searches for $0\nu$DBD because of high energy release
$Q_{\beta\beta}=3.034$ MeV and quite large natural abundance $\delta=9.82\%$. Scintillating bolometers are considered now as one of the most perspective tools in
high-sensitive searches for $0\nu$DBD because of high efficiency (realized in the ``source = detector''
approach) and possibility to distinguish $\beta\beta$ signal from backgrounds ($\alpha$ decays, pile-ups,
etc.) by simultaneous measurement of the heat and light channels \cite{Pir06}.
Recently, several compounds containing Mo have been studied~\cite{Limone}\cite{Lee11}\cite{ZnMoO4}\cite{PbMoO4}, all of them have shown a great ability to identify $\alpha$ interactions from $\beta$/$\gamma$ ones. This feature is extremely important for backgrounds identification and suppression.\newline

It should be mentioned that Li-containing bolometer are interesting target for the investigation of Solar Axions. $^7$Li has a high natural isotopic abundance, about  92\%. Therefore the material can be used as a target to search for hypothetical axions emitted in the solar core in de-excitation of the 478~keV level, which is being populated in one of branches of the pp chain.\newline

In this work we will discuss performances and feature of two Li-containing scintillating bolometer: \LMO and \LEO. The two detectors were operated inside a $^3$He/$^4$He dilution refrigerator in the underground laboratory of Laboratori Nazionali del Gran Sasso (Italy). The detector design is similar to the ones already presented in some works of our group~\cite{Limone}\cite{ZnMoO4}\cite{PbWO4}.

\section{\LMO scintillating bolometer}

\begin{figure}
\centering
\includegraphics[width=8cm,clip]{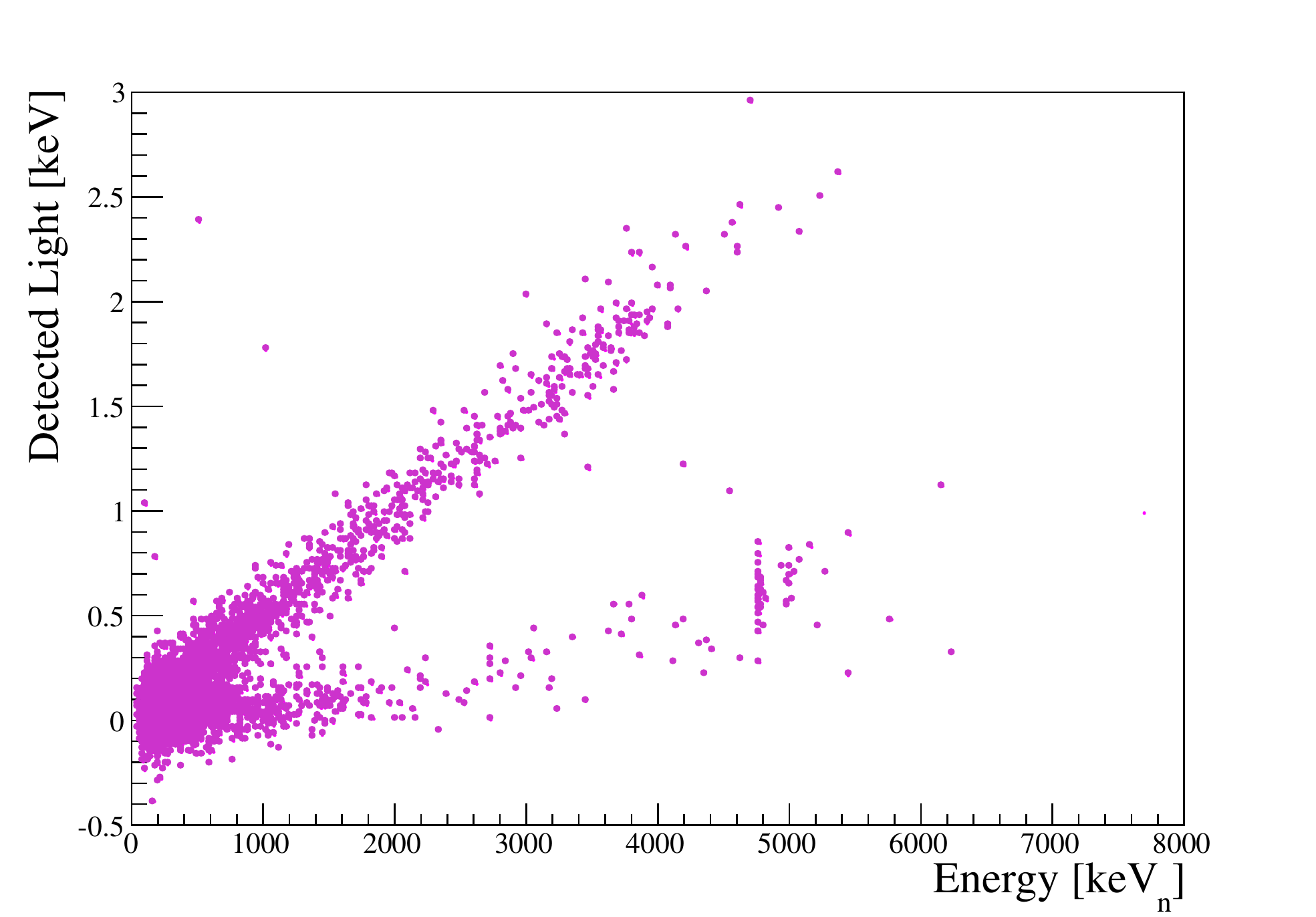}
\caption{Calibration measurement of a \LMO scintillating bolometer with an Am-Be neutron source. In the lower band are visible neutron elastic scatterings in the crystal (from low energy up to 4.78~MeV), thermal and fast neutron absorptions (from 4.78~MeV to higher energies.}
\label{fig:neutron}       
\end{figure}

We have characterized a 33~g \LMO crystal as scintillating bolometer for more than 400~h. Various sources ($\alpha$, $\beta$/$\gamma$ and neutrons) were used for the evaluation of the light yield (LY) of different interacting particles.\newline
\begin{figure}
\centering
\includegraphics[width=8cm,clip]{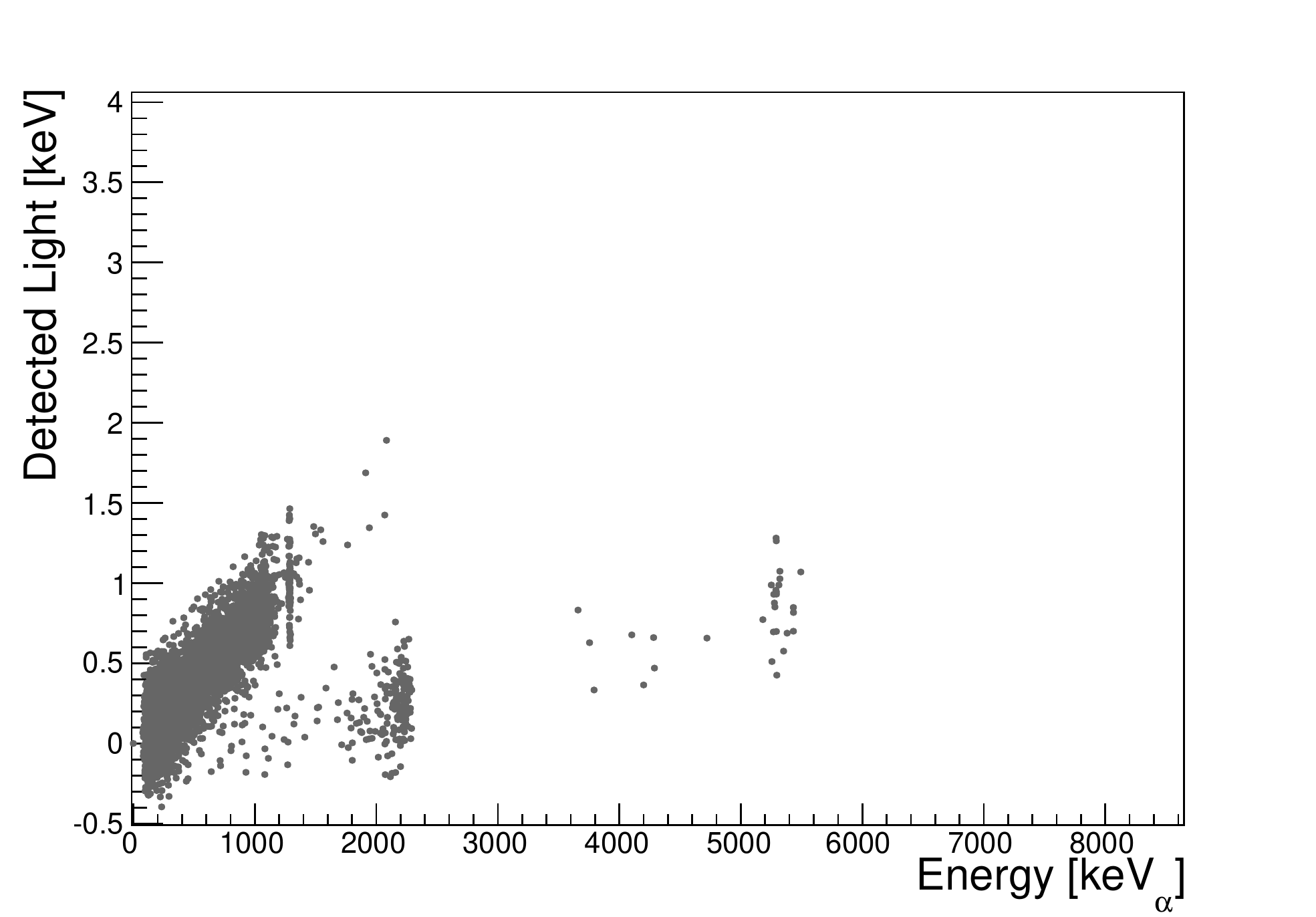}
\caption{Calibration measurement of a \LMO scintillating bolometer with $\beta$/$\gamma$ and $\alpha$ sources. The upper band is ascribed to $\beta$/$\gamma$ events produced by the $^{40}$K source, while in the lower band are visible $\alpha$ interactions in the crystal, namely events produced by a $^{147}$Sm source (at 2.2~MeV) and $^{210}$Po (at 5.4~MeV).}
\label{fig:LMO}       
\end{figure}
The crystal was exposed to a $^{40}$K $\gamma$ source (see Fig.~\ref{fig:LMO}). The computed LY for $\beta$/$\gamma$ interactions is:
\begin{equation}
LY_{\beta / \gamma} = 0.43 \pm 0.01 \; keV/MeV.
\end{equation}
While for the evaluation of the LY for $\alpha$ particles, a $^{147}$Sm source was deposited on the detector set-up, facing the crystal. The computed LY for 2.2~MeV $\alpha$ interaction is 0.127$\pm$0.001~keV/MeV (see Fig.~\ref{fig:LMO}). Given the LY$_{\beta/\gamma}$ and the LY$_{\alpha}$($^{147}$Sm) we can easily compute the quenching factor for $\alpha$ particles with respect to $\beta$/$\gamma$ interaction at 2.2~MeV:
\begin{equation}
QF_{\alpha}(2.2~MeV) = 0.29\pm0.01.
\end{equation}
The Li-based scintillating bolometer was also exposed to a neutron flux of about 200~$n$/s, produced by an Am-Be neutron source. The goal of the measurement was to observe the $^6$Li(n,$\alpha$)$^3$He reaction:
\begin{equation}
\label{reaction}
^6Li \; + \; n \; \rightarrow \; ^3H \; + \; ^4He \; +\; 4.78~MeV.
\end{equation}
which is particularly interesting because it allows to perform neutron spectroscopy studies by means of the neutron energy transfer. In Fig.~\ref{fig:neutron} is shown the acquired energy scatter plot for the calibration measurement with the neutron source.\newline
The detector was also operated in background conditions for more than 300~h. We evaluated the internal contaminations of the crystal by analyzing the $\alpha$ region of the spectrum, due to the more favorable signal-to-noise ratio. In Tab.~\ref{tab:contaminations} are shown the results. 
\begin{table}
\caption{Internal radioactive contaminations for the Li$_2$MoO$_4$ crystal. Limits are computed at 90\% C.L.} 
\begin{center}
\begin{tabular}{lcc}
\hline\noalign{\smallskip}
Chain & Nuclide  & Activity \\ 
            & & [$\mu$Bq/kg] \\
\noalign{\smallskip}\hline\noalign{\smallskip}
$^{232}$Th & $^{232}$Th & $<$~94 \\
\noalign{\smallskip}\hline\noalign{\smallskip}
$^{238}$U & $^{238}$U & $<$~107 \\
 & $^{210}$Pb & 729$\pm$160 \\
\noalign{\smallskip}\hline
\end{tabular}
\label{tab:contaminations} 
\end{center}
\end{table}

\section{\LEO scintillating bolometer}
\begin{figure}
\centering
\includegraphics[width=9cm,clip]{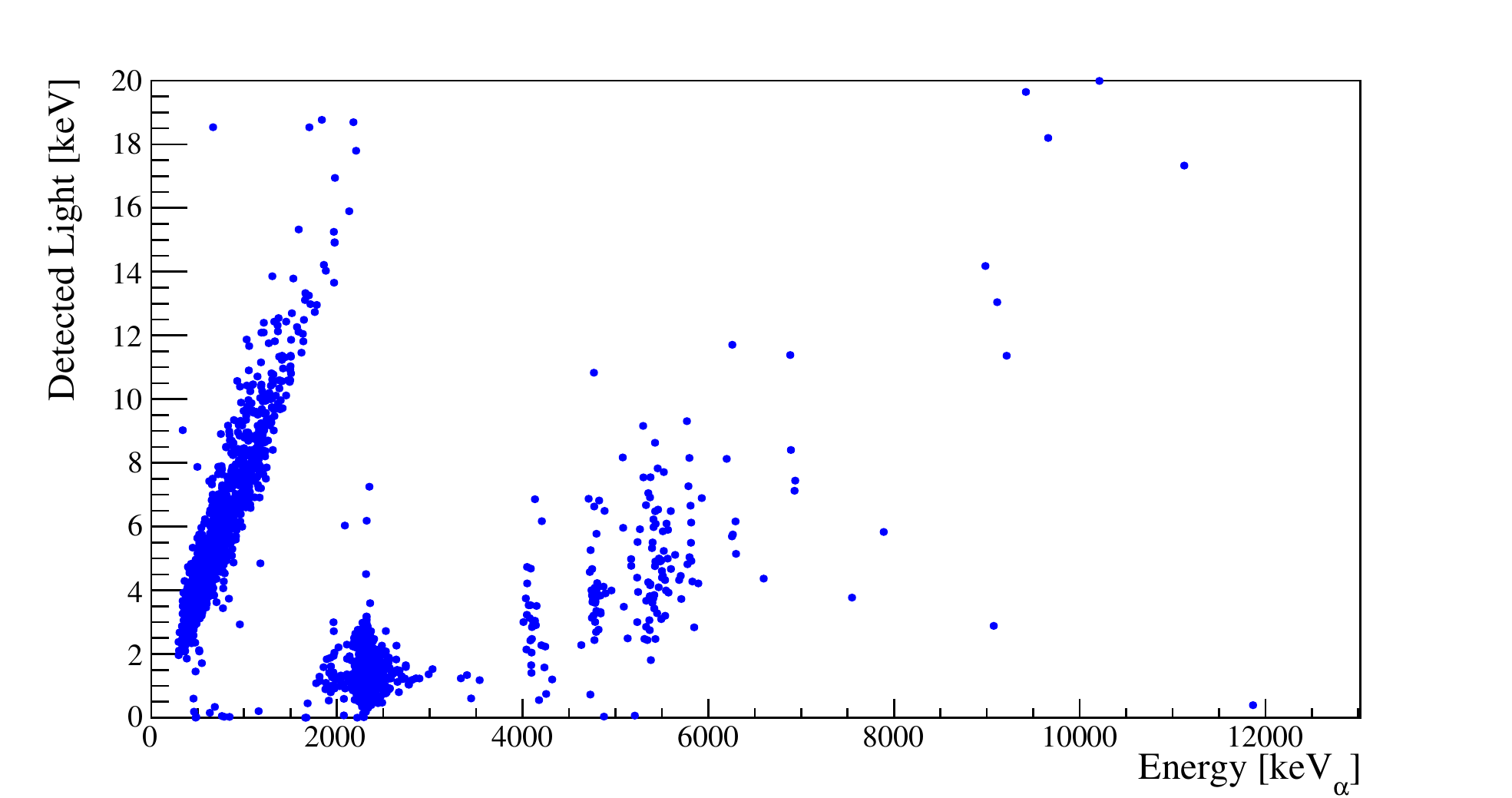}
\caption{Background measurement of a \LEO scintillating bolometer. The upper band is ascribed to $\beta$/$\gamma$ events produced, while in the lower band are visible $\alpha$ interactions in the crystal.}
\label{fig:LEO}       
\end{figure}

We successfully tested for the first time a 6.15~g \LEO as scintillating bolometer for more than 300~h in background conditions. In Fig.~\ref{fig:LEO} is shown the total acquired statistics.
Due to the small crystal mass only a calibration with a $^{137}$Cs $\gamma$ source was performed. The computed LY$_{\beta/\gamma}$ is:
\begin{equation}
LY_{\beta/\gamma} = 6.55\pm0.04~keV/MeV.
\end{equation}
By means of the internal contaminations, namely an intense $^{147}$Sm contamination, we were able to evaluate the LY$_{\alpha}$ and QF$_{\alpha}$:
\begin{equation}
LY_{\alpha}(^{147}Sm) = 0.54\pm0.01~keV/MeV
\end{equation}
\begin{equation}
QF_{\alpha}(^{147}Sm) = 0.08\pm0.01.
\end{equation}
Finally we evaluated the internal contaminations of the crystal using the same method adopted for the \LMO crystal. In Tab.~\ref{tab:contaminationsLEO} the computed values are presented.

The \LEO crystal shows low intrinsic radiopurity level, $^{238}$U/$^{232}$Th decay chain products are observed together with an intense $^{147}$Sm contamination. Given these level of internal contaminations further studies are mandatory for lowering the radiopurity of this interesting compound. 
\begin{table}
\caption{Internal radioactive contaminations for the \LEO crystal. Limits are computed at 90\% C.L.} 
\begin{center}
\begin{tabular}{lcc}
\hline\noalign{\smallskip}
Chain & Nuclide  & Activity \\ 
            & & [mBq/kg] \\
\noalign{\smallskip}\hline\noalign{\smallskip}
$^{232}$Th & $^{232}$Th & 3.5$\pm$0.8 \\
\noalign{\smallskip}\hline\noalign{\smallskip}
$^{238}$U & $^{238}$U & $<$0.3 \\
 & $^{226}$Ra & 2.9$\pm$0.7 \\
 & $^{210}$Po & 6.2$\pm$0.9 \\
\noalign{\smallskip}\hline\noalign{\smallskip}
$^{147}$Sm & $^{147}$Sm & 454$\pm$9 \\
\noalign{\smallskip}\hline
\end{tabular}
\label{tab:contaminationsLEO} 
\end{center}
\end{table}

\section{Conclusion}
Two Li-based scintillating bolometers were assembled and successfully tested. The compounds shows interesting features in terms of particle discrimination as shown in Fig.~\ref{fig:neutron} and Fig.~\ref{fig:LEO}, thanks to the double read-out (heat vs. light). The tested crystal have small masses, and further R\&D and studies are needed in order to adopt these kind of materials as targets in multi-ton DM and DBD experiments.

\section*{Acknowledgements}
This work was made in the frame of the LUCIFER experiment, funded by the European Research Council under the European Union's Seventh Framework Program (FP7/2007-2013)/ERC grant agreement no. 247115.

%
%

\end{document}